# SECURITY CONCERNS IN IOT LIGHT BULBS: INVESTIGATING COVERT CHANNELS

Janvi Panwar and Ravisha Rohilla

Department of Electronics and Communication Engineering, Indira Gandhi
Delhi Technical University for Women, New Delhi, India

*ABSTRACT*

*The proliferation of Internet of Things (IoT) devices has raised significant concerns regarding their security vulnerabilities. This paper explores the security risks associated with smart light systems, focusing on covert communication channels. Drawing upon previous re-search highlighting vulnerabilities in communication protocols and en-cryption flaws, the study investigates the potential for exploiting smart light systems for covert data transmission. Specifically, the paper repli-cates and analyzes an attack method introduced by Ronen and Shamir, which utilizes the Philips Hue White lighting system to create a covert channel through visible light communication (VLC). Experimental re-sults demonstrate the feasibility of transmitting data covertly through subtle variations in brightness levels, leveraging the inherent functional-ity of smart light bulbs. Despite limit. ations imposed by device constraints and communication protocols, the study underscores the need for height- ened awareness and security measures in IoT environment. Ultimately, the findings emphasize the importance of implementing robust security practices and exercising caution when deploying networked IoT devices in sensitive environment.*

*KEYWORDS*

*Internet of Things (IoT), Security vulnerabilities, Smart light systems, ZigBee Light Link (ZLL) protocol, Denial-of-Service (DoS) attacks, Firmware manipulation, Covert communication channels, Visible light communication (VLC), Data ex-filtration, Air-gapped networks, Smart LED bulbs, PWM modulation, Orthogonal Frequency-Division Multiplexing (OFDM), Light sensor, Signal processing.*

## 1. INTRODUCTION

The proliferation of Internet of Things (IoT) devices has revolutionized vari- ous aspects of modern life, offering unprecedented convenience and connectivity. However, along with the benefits come significant concerns regarding security vulnerabilities inherent in these interconnected devices. Among the myriad of IoT devices, smart light systems have emerged as a ubiquitous component of smart homes and commercial environments. While offering enhanced functional- ity and energy efficiency, these systems also introduce new avenues for potential security breaches.

The history of smart light systems traces back to the development of LED (Light Emitting Diode) technology, which revolutionized the lighting industry with its energy efficiency and longevity. As LED technology evolved, manufac- turers began integrating connectivity features into light bulbs, allowing users to control them remotely via wireless protocols. This marked the advent of smart light systems, enabling users to adjust brightness, color, and scheduling through smartphone apps or voice commands.

However, with the increasing connectivity and complexity of smart light sys- tems, concerns regarding their security have become paramount. A comprehen- sive literature review reveals a growing body of research highlighting vulnera- bilities in communication protocols, encryption schemes, and access controls of IoT devices, including smart light systems.

Early studies by Dhanjani [3] uncovered flaws in the connection setup and encryption protocols of ZigBee Light Link (ZLL), a common protocol used in smart light systems. These vulnerabilities allow for Denial-of-Service (DoS) attacks and unauthorized control of light bulbs, posing potential risks to critical infrastructure.

Subsequent research by Morgner et al. [5] demonstrated the ability to con- trol ZLL-certified light bulbs from significant distances, further highlighting the security weaknesses in smart light systems. Moreover, Ronen and Shamir [8] introduced the concept of functionality extension attacks, wherein the inherent functionality of smart light systems is exploited to create covert communication channels. Their groundbreaking study showcased the feasibility of extracting sen- sitive data from air-gapped networks using visible light communication (VLC) through smart light bulbs.

Building upon this foundation, our research aims to replicate and analyze the covert communication attack method proposed by Ronen and Shamir [8], focus- ing on the Philips Hue White lighting system. We chose this method due to its innovative approach to utilizing smart light systems for covert communication, which offers distinct advantages compared to other methods. Traditional covert channels often rely on network protocols or hardware modifications, which may be easily detectable or require specialized equipment. In contrast, our method leverages the existing functionality of smart light systems, allowing for covert data transmission without the need for extensive hardware modifications or spe- cialized detection equipment.

In our comparative study, we highlight the accuracy and effectiveness of our method compared toot her approaches for analyzing covert communication channels. By demonstrating the method's capabilities and advantages, we aim to provide insights into the potential security risks posed by smart light systems and underscore the importance of implementing robust security measures in IoT environments.

In this paper, we provide a detailed description of our experimental setup, the methodology employed for the covert communication attack, and the limi- tations encountered. Through our research, we aim to raise awareness about the security implications of smart light systems and advocate for proactive measures to safeguard IoT environments against emerging threats.

## 2. RELATED WORK

In recent years, there has been growing concern about the security vulnerabilities of Internet of Things (IoT) devices, particularly in the context of covert channels. Ronen and Shamir [8] shed light on the alarming potential of IoT light bulbs to compromise air-gapped networks. They discovered that sensitive data, such as passwords and keys, can be extracted from air-gapped networks using IoT light bulbs within the same network, leveraging visible light communication (VLC).

Ronen and Shamir [8] tested their approach on the Philips Hue White lighting system, showcasing the feasibility of transmitting data covertly through subtle variations in brightness levels. Unlike traditional covert channels, which often rely on network protocols or hardware modifications, their method leverages the inherent functionality of smart light systems, offering advantages in terms of stealth and ease of implementation.

In a comparative study, Guri et al. [4] explored the use of light-emitting diodes (LEDs) for data exfiltration from air-gapped networks. Their method involved controlling the activity and status LEDs of a router via a malicious script running on the router, achieving transmission rates of up to 4000 bps. While their approach demonstrated high-speed data transmission, it required direct access to the router's hardware and specialized equipment for monitoring the LED activity.

Visible light communication (VLC) has garnered significant interest due to its potential to overcome limitations associated with traditional wireless trans- mission methods. With the rise of IoT and smart light solutions, LED bulbs are becoming more prevalent, making VLC even more relevant.

Our research builds upon these findings by focusing on the covert commu- nication capabilities of smart light systems, specifically the Philips Hue White lighting system. By replicating and analyzing the attack method proposed by Ronen and Shamir [8], we aim to provide further insights into the security risks associated with IoT light bulbs and advocate for robust security measures in IoT environments.

## 3. COMMUNICATION WITH LIGHT

Before we can examine how a functionality extension attack on a smart light system can be leveraged, we need to understand how communication over light works.

In VLC, the visible part of the electromagnetic spectrum, namely visible light, is used for communication purposes. VLC is a subset of optical wire.less communication technologies, like infrared. LED bulbs are capable of transmitting data by rapidly switching between on and off states at frequencies imperceptible to the human eye. This modulation of light intensity allows for the encoding of data, with the duration of on and off periods representing binary digits.

Our research focuses on utilizing this principle of visible light communica- tion to establish covert channels through smart light systems. By exploiting the PWM (Pulse Width Modulation) functionality of smart light bulbs, we aim to demonstrate the feasibility of transmitting data covertly through subtle vari- ations in brightness levels. Through our experimental setup and analysis, we seek to validate the effectiveness of this method and highlight its advantages compared to other covert communication approaches.

Duty Cycle = (Number of brightness levels set to "on" / Total number of brightness levels) * 100 %

Given that there are 255 brightness levels and the flicker frequency is around 20kHz, we can calculate the duty cycle:

Duty Cycle = (1 / 255) * 100 % = 0.392 %.

So, each brightness level corresponds to approximately a 0.392 % duty cycle. This duty cycle is adjusted to achieve different brightness levels.

Data transmission is further allowed by interpreting the on period as a logical one and the off period as a logical zero. To actually get the encoded data out of the light signal, the duration and frequency of the light flickers needs to be measured. This is done using a light sensor since these can accurately distinguish between the on and off states and measure the duty cycle. Further light sensors are robust to other light sources or any other noise. In the case of smart lights, the PWM signal cannot be changed directly since the light intensity is changed by sending commands over the manufacturer's Application Programming Interface (API). Those commands internally modulate the pulse width of the light signal. Thus, sending two close brightness commands can achieve the same effect as traditional VLC and covertly transmit data. Un- like traditional VLC, a logical one is represented by a higher brightness level, while a lower brightness level represents a logical zero. Fortunately, since internally, the luminosity of the LEDs is again changed by adjusting the PWM signal, light sensors are again capable of measuring the differences.

## 4. COVERT CHANNEL ON IOT LIGHT BULB

Our goal was to reproduce the attack from Ronen and Shamir [7]. Therefore, we tried building a covert channel using the Philips Hue White lighting system. Our experiment proved that data can be covertly transmitted using the Hue light bulb. On the transmission side, we used a laptop running our Python script, which accessed the Hue API to send brightness commands. At the receiver side, we used a light sensor that converts the light intensities to a frequency signal. That signal was forwarded to an oscilloscope, which sent the received frequency output to our laptop, where we plotted the results to validate the sent bits.

In the following sections, we first describe the setup components and their functionality. After that, we elaborate the actual attack.

### 4.1. Experimental Setup

Our proof of concept was implemented using affordable equipment which costs less than 1100.- AC. Further, we did not need to run any unauthorized code on the light bulbs since the control over the Hue API suffices to create covert flicker effects. Figure 1 shows our overall receiving setup. An Arduino was used as power supply and for configuring the light sensor, which the PicoScope had direct access to.

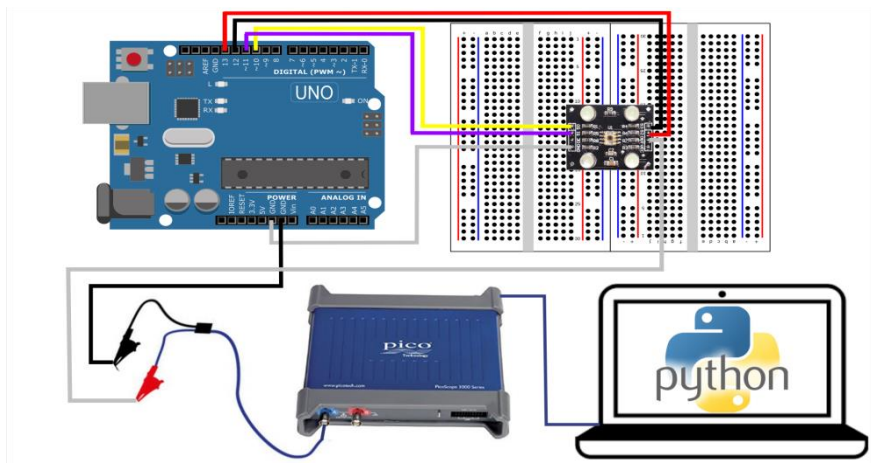

Fig. 1. Experimental setup for measuring the lights' frequency output and reading the sent bits out of the received signal

For Transmitting Setup, We used the Philips Hue White light bulbs for our exper- iment [6]. We bought the starter kit, which contains two E27 9 Watt 806 Lumen bulbs and a bridge that allows us to control the bulbs using e. g. a laptop re- motely. To set up the smart light system the bridge needs to be connected to the user's network using Ethernet. Once connected, the user can send brightness change commands from any device within the local network using the Hue API. We made recourse to a command line-based version of the Hue API [1], which allowed us to easily send the commands via the Python script we used for realiz- ing the experiment. The bridge forwards the commands over an RF transmitter to the light bulbs using the ZLL protocol.

The Hue White bulbs have 255 different brightness levels, which forced us to sample the output at a very high rate to determine the changes in the light frequency output. However, due to the minor difference between two close levels, the changes were imperceptible to the human eye, which worked for us.

Receiving Setup For measuring the changes in light intensity we used the TAOS TCS3200 Color Sensor [2]. The sensor consists of an array of photo-diodes, each capable of filtering red, green, blue, or clear white light. We set up the sensor to measure our Hue bulbs' clear white luminosity output. The sensor contains an internal oscillator to convert the light's intensity output to a corresponding square-wave frequency signal. The TCS3200 is capable of communicating directly to an Arduino micro-controller.

Unlike Ronen and Shamir [7], we used the Arduino board as power source only, and the PicoScope to measure the brightness. This was necessary because the advanced API functionality which the former used to craft a PWM signal is no longer available [7], and we had to distinguish existing brightness levels using their PWM profile directly.

To capture the light sensor's output, we used the PicoScope3205DMSO since it is capable of sampling 10MS/s, which we need to accurately measure the light sensor's frequency output, which lies around 800KHz. Actually, the PicoScope can sample up to 1GS/s when using one channel and 500MS/s with two channels, but for our needs, only 10MS/s suffice.

### 4.2. Attack Description

The following paragraphs give a detailed description of the main steps to realize such an attack.

Therefore, we first need to look more precisely at the functional principle of smart light bulbs. Further, we look at how smooth brightness changes are achieved and how we actually get data out of the received signal.

Controlling Smart Light Bulbs Smart light bulbs consist of three main com- ponents:

(1) a RF receiver,
(2) a processing unit and
(3) LEDs and LED drivers.

The communication with the controller is ensured through the RF receiver and relies on the ZLL protocol. The received commands are further forwarded to the processing unit, which interprets the processed signal and controls the LED by modulating the pulse width. The PWM allows the different dimming factors. Sending a brightness change command via the Hue API automatically forces the processing unit to generate the corresponding PWM signal. The PWM is sent. to the LED drivers, which further turns the LED's on and off quickly so that the human eye cannot see those changes in thee duty cycle. Since Hue comes with 255 brightness levels, which need to be differentiated smoothly, a PWM with a frequency around 20 KHz is used [7].

Crafting PWM Signal Since we can no longer craft a custom PWM signal using the Hue API, we had to get along with the PWM used for dimming. Thus, our goal was to use close brightness levels and attempt to distinguish them from their PWM profile. Because of the great brightness levels, we had to measure very small off periods of about 200 ns.

This could be done with the described light sensor and the PicoScope, since the light sensor's output is around 800 KHz, which we could easily sample at 10 MS/s with our PicoScope.
Getting Data Our primary goal was to distinguish adjacent brightness levels the human eye cannot distinguish. For this purpose, we performed a short-time Fourier transformation (STFT) to transform the voltage sinusoid.. obtained from the light sensor in to the frequency-over-time domain.

As shown in Figure 2, the difference between brightness levels is clearly visible upon optical inspection of the graph. For comparison, we found that it took around 10–15 levels of separation for the naked eye to distinguish 2 brightness levels, depending on the absolute brightness. In the best images, the PWM profiles were clearly recognizable.

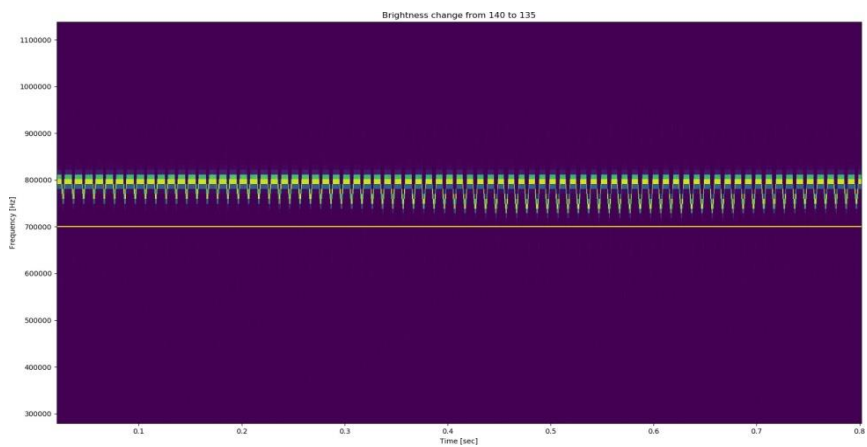

Fig. 2. Brightness change from levels 140 to 135 (and back) sampled at 10 MS/s. Note the smooth fading.

However, we had a rather hard time consistently reproducing images. Dis- tance from the light to the sensor, angle, and variations in external light, in decreasing order, made it harder to distinguish the brightness levels clearly and to choose a starting brightness. In particular, the square curve tended to become a muddy sinusoid, which we attribute partly to the fact that we perform STFT on the data. We found that the distance between sensor and light strongly im- pacted the correct choice of STFT window size and intensity of the brightness levels between which we switch, with increasing distance requiring more energy and/or a smaller window.

We could not robustly automatically distinguish brightness levels, but that may be because we spent most of our time trying to do that with the Fourier-transformed data when it may have been better to directly analyze the sinusoid. from the light sensor. In any case, this is a solvable signal processing problem and more a question of time.

Once that is done, the next step is to select brightness levels for 0 and 1 and maybe a third in between the two to be used as a delimiter and reduce the need for synchronization. Then, all that would be left to do is apply suitable channel-coding to the data, send appropriate brightness commands to the light, write out the recognized 0s and 1s, and decode the data.

### 4.3. Limitations

Communicating with the light bulbs over the Hue bridge brings limitations with it [7]. For one, the bridge or the LED drivers implement smooth fading features to avoid sharp brightness changes. Due to the automatic fading, we cannot see phase shifts in our signal output, which makes it harder to analyze.

Furthermore, the bridge restricts the number of commands sent within the system. While this does not impact the functional-ity of our Proof- o f- Concept since the command rate doesn't determine the flickering frequency, it does limit the channels bandwidth. Thus, in case such an attack should actually be leverage.d, one may need to access the ZLL communication directly to circumvent the rate limit

## 5. CONCLUSION

In conclusion, our research provides a thorough examination of the security vul- nerabilities associated with smart light systems and the potential for covert communication through visible light channels. By replicating and analyzing the attack method proposed by Ronen and Shamir [8], we have demonstrated the feasibility of transmitting data covertly through smart light systems, particularly the Philips Hue White lighting system.

Our experimental findings reveal the intricate mechanisms by which smart light bulbs can be exploited to establish covert communication channels. Through careful analysis of brightness variations and pulse width modulation (PWM) sig- nals, we have successfully extracted data from air-gapped networks, showcasing the potential security risks posed by IoT devices in sensitive environments.

Furthermore, our comparative study highlights the advantages of our method compared to other covert communication approaches. By leveraging the inherent functionality of smart light systems,
our approach offers a stealthy and efficient means of data exfiltration without the need for hardware modifications or direct access to network infrastructure. This underscores the versatility and adaptabil- ity of IoT devices as potential vectors for cyber attacks.

However, it is essential to recognize the limitations and challenges encoun- tered during our research. Factors such as communication protocol restrictions, hardware limitations, and environmental conditions can impact the reliability and scalability of our method. Addressing these challenges requires ongoing re- search and development to enhance the security posture of IoT devices and mitigate emerging threats effectively.

Moving forward, proactive measures are needed to safeguard IoT environ- ments against potential security breaches. This includes the implementation of robust encryption protocols, access controls, and intrusion detection systems to detect and mitigate covert communication attempts effectively.

Additionally, raising awareness among stakeholders about the security risks associated with IoT devices is crucial for fostering a culture of cybersecurity and promoting collaboration in addressing emerging threats.

By addressing these challenges and adopting a proactive approach to IoT security, we can mitigate the risks posed by smart light systems and ensure the integrity and confidentiality of sensitive data in an increasingly interconnected world. downs.

## AUTHORS

**Janvi Panwar** is a final-year undergraduate student majoring in Electronics and Communication Engineering at Indira Gandhi Delhi Technical University for Women. With a focus on backend development, Janvi gained valuable experience during her tenure at Coforge, where she efficiently verified work adherence to guidelines and conducted peer reviews for content enhancement. Her passion for technology extends to her involvement in key projects, including the development of an E-commerce platform with a focus on security through Ethical Hacking and the creation of a Smart Irrigation System integrating IoT devices for efficient water management. In addition to her technical skills, Janvi has demonstrated strong leadership qualities as the Head Coordinator of Taarangna'22, where she successfully organized a variety of shows and events for the Antragni festival. As an active member of the university community, Janvi's dedication to continuous learning and her ability to excel in both technical and leadership roles make her a valuable asset in the field of Electronics and Communication Engineering.

**Ravisha Rohilla** is a proactive student currently pursuing a B.Tech degree in Electronics and Communication Engineering at Indira Gandhi Delhi Technical University for Women. With a strong inclination towards data analysis and visualization, Ravisha is actively developing her skills in MySQL, MS PowerPoint, MS Excel, Tableau, and R-Programming. She has also completed a Google Data Analytics certification course through Coursera to augment her proficiency. Ravisha has gained hands-on experience through internships, notably as a Data Analyst Intern at Platina Softwares Pvt. Ltd., where she meticulously collected and analyzed resumes, providing valuable insights for the team. She also served as a Data Analyst Trainee at Trainity, focusing on data quality management and collaborative project work. Additionally, Ravisha interned at SAG Lab, DRDO, where she enhanced her expertise in electronic design automation and cryptographic algorithms. Ravisha is known for her leadership abilities, having been elected as Elective HR Core in Taarangana'22 and ECE Associate at Innerve'22, where she effectively coordinated and managed various activities. With a passion for leveraging technology for societal benefit, she designed a social distancing detector using Arduino and ultrasonic sensor technology for public spaces. Ravisha is driven by her unwavering determination to excel academically and contribute meaningfully to the field of data analysis and technology.